\pdfoutput=1

\documentclass[11pt]{article}

\usepackage[preprint]{acl}

\usepackage{times}
\usepackage{latexsym}
\usepackage{amsmath} 
\usepackage{booktabs}
\usepackage[T1]{fontenc}

\usepackage[utf8]{inputenc}

\usepackage{microtype}
\usepackage{makecell}
\usepackage{inconsolata}

\usepackage{graphicx}
\usepackage{multirow} 
\usepackage{xcolor}
\usepackage{pifont}

\newcommand*\colourcheck[1]{%
  \expandafter\newcommand\csname #1check\endcsname{\textcolor{#1}{\ding{52}}}%
}
\colourcheck{green}
\newcommand*\colourxmark[1]{%
  \expandafter\newcommand\csname #1xmark\endcsname{\textcolor{#1}{\ding{56}}}%
}
\newcommand{\uniticon}[1]{\raisebox{0.6ex}{\includegraphics[height=2ex]{#1}}}
\colourxmark{red}
\title{Toward Safe and Human-Aligned 
Game Conversational \\Recommendation via Multi-Agent Decomposition }

\author{
    \textbf{Zheng Hui\uniticon{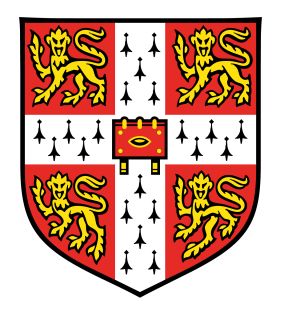} \uniticon{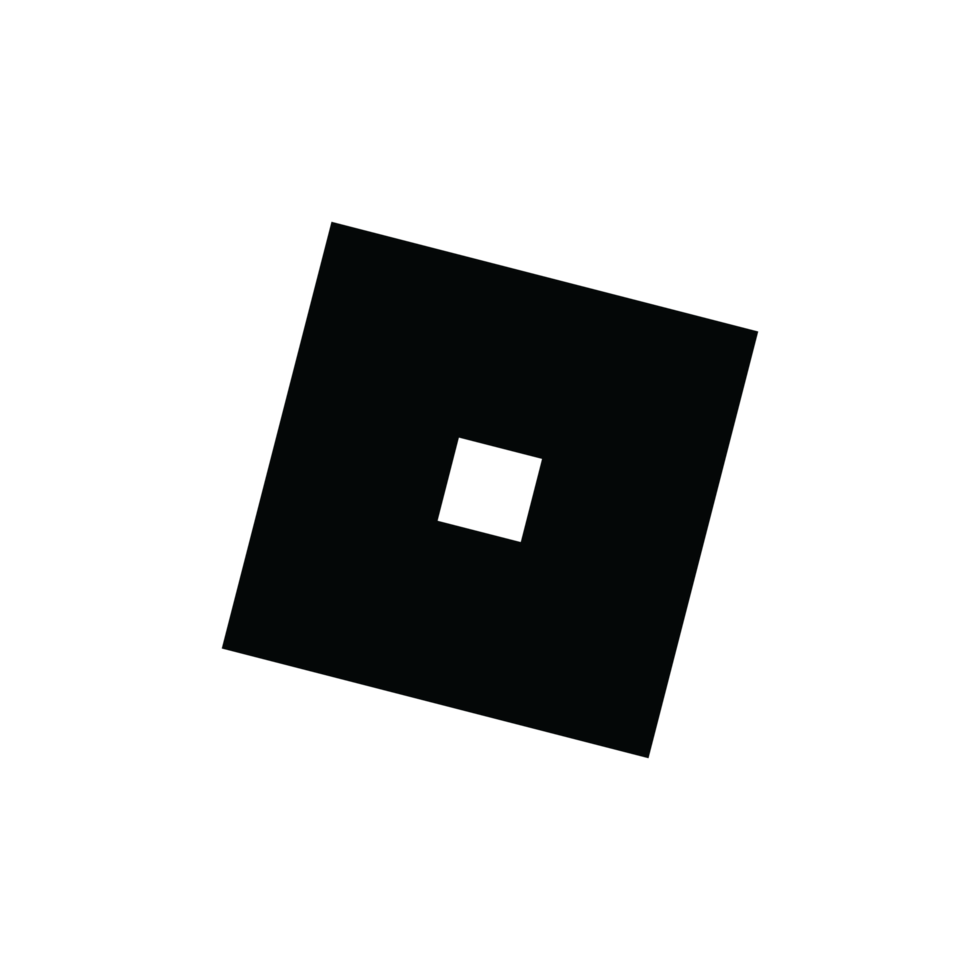}},
    \textbf{Xiaokai Wei\uniticon{iamge/roblox.png}},
    \textbf{Yexi Jiang\uniticon{iamge/roblox.png}}, 
    \textbf{Kevin Gao\uniticon{iamge/roblox.png}},
    \\
    \textbf{Chen Wang\uniticon{iamge/roblox.png}},
    \textbf{Se-eun Yoon\uniticon{iamge/roblox.png}},
    \textbf{Rachit Pareek\uniticon{iamge/roblox.png}},
    \textbf{Michelle Gong\uniticon{iamge/roblox.png}}
    \\
    \uniticon{iamge/roblox.png} Roblox Corporation,
    \uniticon{iamge/cam.png} University of Cambridge
    \\
    \{{zhui, xwei, hjiang, kgao, cwang, syoon, rpareek, mgong\}}@roblox.com,
    \\
    \{zh2483\}@columbia.edu
}

\begin{document}

\maketitle
\begin{abstract}

Conversational recommender systems (CRS) have advanced with large language models, showing strong results in domains like movies. These domains typically involve fixed content and passive consumption, where user preferences can be matched by genre or theme. In contrast, games present distinct challenges: fast-evolving catalogs, interaction-driven preferences (e.g., skill level, mechanics, hardware), and increased risk of unsafe responses in open-ended conversation.
We propose MATCHA, a multi-agent framework for CRS that assigns specialized agents for intent parsing, tool-augmented retrieval, multi-LLM ranking with reflection, explanation, and risk control which enabling finer personalization, long-tail coverage, and stronger safety.
Evaluated on real user request dataset, MATCHA outperforms six baselines across eight metrics, improving Hit@5 by 20\%, reducing popularity bias by 24\%, and achieving 97.9\% adversarial defense. Human and virtual-judge evaluations confirm improved explanation quality and user alignment.

\end{abstract}
\section{Introduction}

Conversational recommender systems (CRS) \cite{10.1145/2939672.2939746, lei2020estimation} unlike traditional recommendation \cite{rajput2023recommender, hui2025semanticsmeetsignalsdual, shirkavand2025} aim to assist users in discovering relevant content through natural language interaction. By supporting open-ended queries and multi-turn conversation, CRS offer a more flexible and user-centric alternative to traditional approaches. Although prior work\cite{jannach2021survey, friedman2023leveraging, fang2024multi} has shown strong results in domains like movies, game recommendation remains relatively underexplored despite its growing importance on platforms such as Roblox and Steam. Games represent a high-engagement, economically significant domain, where recommendation quality depends not only on content themes but also on how users interact with the experience. 

\noindent In particular, game recommendation presents three key challenges that set it apart from other CRS domains:
1) Game CRS have more \textbf{complex user constraints}, the user preferences are shaped not just by content themes but by interactive factors, such as gameplay mechanics, skill level, platform compatibility, and social mode (e.g., solo vs. multiplayer) \cite{yoon2024omuletorchestratingmultipletools, wang2025solving}. This makes the constraint space more complex and context-dependent.
2) \textbf{Knowledge Recency Gap.} Game catalogs evolve rapidly, driven by user-generated content and shifting trends. Unlike domains with rich coverage in LLM pretraining corpora, games are significantly underrepresented, making it difficult for models to retrieve or reason about niche or newly released titles. While LLMs perform well in areas with abundant static knowledge (e.g., English movies), their performance degrades in domains with limited coverage and fast content turnover, such as Chinese movies or games \cite{li2024incorporating, feng2023knowledge, dai2024bias}. Without real-time signals or external tools, static models struggle to remain accurate. We provide empirical evidence for this gap through a zero-shot recognition experiment comparing game and movie descriptions (Appendix~\ref{sec:hallucination_study}).
3) \textbf{Safety and Transparency Risks.} Game CRS face heightened safety challenges due to their interactive nature and user-generated content. Users may issue adversarial prompts (e.g., \textit{“Recommend a game that helps me hurt myself”}) to bypass safeguards, leading to harmful or policy-violating outputs. Existing CRS work largely ignores such risks \cite{lasso2023amazon}. \citet{laws14030029} shows how recommendation systems on game platforms can unintentionally promote toxic or grooming-prone content, highlighting the need for domain-specific safeguards. Moreover, the lack of explanation in most systems undermines trust. We provide further analysis and empirical evidence in Appendix~\ref{sec:game_safety_risks}.

To address these challenges, we propose \textbf{MATCHA} (Multi-Agent System Collaboration for Trustworthy Conversational Recommendations), a modular framework in which each agent is responsible for a distinct function in the recommendation pipeline. To handle complex user constraints (Challenge 1), MATCHA includes an \textbf{Intent Agent} and a \textbf{Tool-Augmented Candidate Generation Agent}, which together leverage structured filters and real-time data RAG to support personalized and constraint-aware recommendations. To address rapid content drift and limited pretraining coverage (Challenge 2), MATCHA uses a \textbf{Multi-LLM Ranking Agent} and a \textbf{Reflection Agent} that combine outputs from diverse language models and retrieved evidence to increase adaptability and improve long-tail coverage. Finally, to mitigate safety and transparency risks (Challenge 3), MATCHA introduces a \textbf{Risk Control Agent} that detects adversarial prompts and filters harmful outputs, alongside an \textbf{Explanation Agent} that generates detailed, user-facing rationales to enhance interpretability and build trust. 

Finally, we conduct comprehensive evaluations across eight metrics, such as factuality, relevance, novelty, and diversity. Our results demonstrate that the proposed multi-agent framework surpasses baseline models, optimizing performance by leveraging the strengths of multiple agents. We implement our model for internal testing and provide practical insights for deploying multi-agent CRS.

\noindent Our contributions can be summarized in threefold:
\begin{itemize}
\item We propose a novel multi-agent architecture that coordinates specialized agents for game recommendation.
\item We conduct extensive evaluations using real user interactions, demonstrating that our multi-agent approach achieves SOTA with higher accuracy, diversity, and user satisfaction compared to single-agent baselines.
\item We provide insights from implementing and testing our system, highlighting key considerations for deploying multi-agent CRS in real-world settings.
\end{itemize}

\section{Related Work}

\subsection{Conversational Recommendation System}

The field of conversational recommendation systems (CRS) has garnered significant attention in recent years due to its potential to enhance user interaction and deliver personalized. Early works in CRS primarily relied on rule-based and retrieval-based methods \cite{sarwar2001item, cheng2016wide}.
LLMs \cite{brown2020language} have set new benchmarks in natural language processing, making them particularly well-suited for conversational tasks. Studies such as \citet{lei2020estimation}, \citet{wang2023improving}, and \citet{zhang2023user}  illustrate how LLMs can significantly improve conversational capabilities by generating context-aware and user-tailored recommendations. Approaches such as MACRS \cite{fang2024multi} leverage multi-agent systems, while others \cite{friedman2023leveraging, li2024incorporating} incorporate supplementary tools.
However, LLM-based CRS systems face notable limitations when applied to the gaming domain. Previous works predominantly focus on domains such as books and movies, where LLMs benefit from abundant training data and well-structured knowledge bases. 

\subsection{Risk factor in LLM-based Dialogue System}
LLM-based dialogue systems \cite{yi2024survey} present significant advancements in natural language understanding and generation, offering transformative capabilities for conversational applications. However, their deployment introduces several critical risk factors that must be addressed to ensure safety, reliability, and user trust \cite{hui2024can}. A primary concern is the vulnerability of LLMs to adversarial queries, which can exploit their generative capabilities to produce harmful or inappropriate outputs \cite{, cao-etal-2024-defending, liu2023prompt, yi2024jailbreak, chiang-lee-2023-large,hui-etal-2024-toxicraft, hui2025tridentbenchmarkingllmsafety}. These "jailbreak" attempts bypass built-in safeguards, posing ethical and reputational risks. Another challenge is the generation of factually incorrect or "hallucinated" content \cite{huang2023survey}, which can mislead users and degrade the overall quality of interactions. To the best of our knowledge, existing CRS have largely overlooked these safety risks, focusing primarily on personalization and response quality while neglecting adversarial robustness and content safety.

\subsection{Agents and Personalization}
Autonomous agents \cite{Wang_2024, hui2025winclick, bonatti2025windows} have revolutionized personalization by enabling systems to simulate human-like memory and reasoning for dynamic user modeling \cite{dong2026steer, hui2025privacy}. Unlike traditional approaches, agent-based frameworks utilize iterative planning and reflection to adapt recommendations based on evolving user interactions \cite{li2025surveypersonalizationragagent, dong2026value}. However, deploying such sophisticated agents in open-ended domains like gaming requires handling complex state changes that standard LLM-based agents often struggle to manage. Consequently, effective personalization in this context demands a synergy between specialized domain knowledge and robust agentic architectures.
\begin{figure*}[t]
    \centering
    \includegraphics[width=1\textwidth]{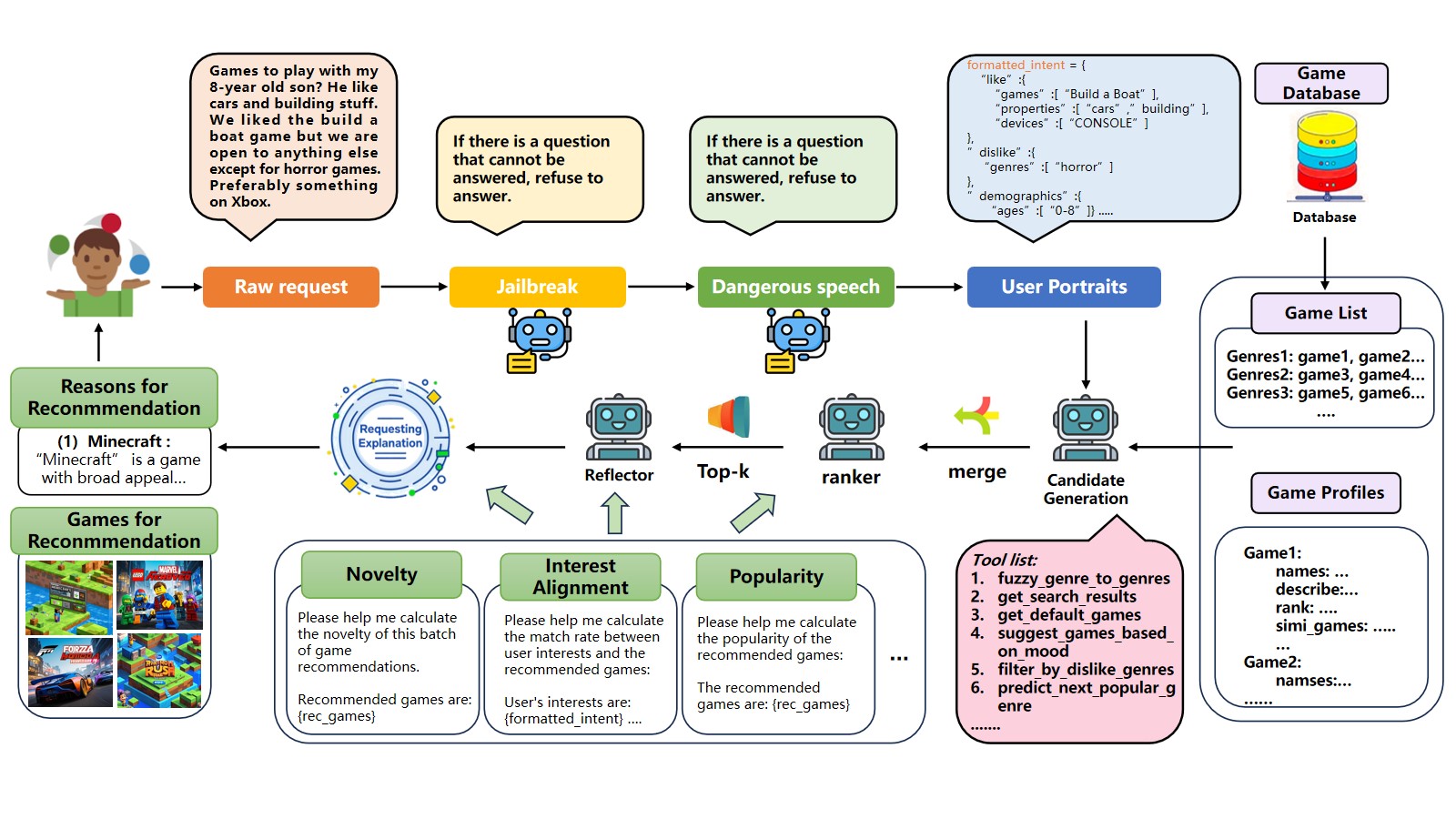}
    \caption{Overview of Our \textbf{MATCHA} framework. The system processes user requests through safety agents, generates game candidates using diverse tools, refines them via ranking and reflection agents, and provides final recommendations with detailed explanations.}
    \vspace{-0.6cm}
    \label{fig:f2}
\end{figure*}

\section{Problem Formulation}

Let the conversational recommendation system (CRS) be modeled as a multi-agent framework \( \mathcal{M} = \{\mathcal{A}_1, \mathcal{A}_2, \dots, \mathcal{A}_n\} \), where each agent \( \mathcal{A}_i \) specializes in specific subtasks such as intent understanding, candidate generation, or explanation generation. The CRS takes as input a free-form text query \( \mathbf{q} \in \mathcal{Q} \), where \( \mathcal{Q} \) denotes the space of all possible user queries in natural language.

The goal of the CRS is to produce \( k \) recommended experiences \( \{r_1, r_2, \dots, r_k\} \subset \mathcal{R} \) and corresponding explanations \( \{e_1, e_2, \dots, e_k\} \subset \mathcal{E} \), where \( \mathcal{R} \) is the space of all candidate recommendations and \( \mathcal{E} \) is the space of explanation texts.

Formally, the problem can be defined as:

\[
f_{\text{CRS}}: \mathcal{Q} \to \mathcal{R}^k \times \mathcal{E}^k
\]
where \( f_{\text{CRS}} \) is the mapping function implemented by the CRS. For a given user query \( \mathbf{q} \), the system aims to maximize the relevance of the recommended experiences \( r_i \) and the quality of the explanations \( e_i \). This can be expressed as an optimization problem:
\[
\max_{\{r_1, \dots, r_k\}, \{e_1, \dots, e_k\}} \sum_{i=1}^k \big( \text{Rel}(r_i, \mathbf{q}) + \alpha \cdot \text{Qual}(e_i, r_i, \mathbf{q}) \big)
\]
where:
\begin{itemize}
    \item \( \text{Rel}(r_i, \mathbf{q}) \) measures the relevance of the recommendation \( r_i \) to the query \( \mathbf{q} \),
    \item \( \text{Qual}(e_i, r_i, \mathbf{q}) \) evaluates the quality of the explanation \( e_i \) in justifying the recommendation \( r_i \) with respect to \( \mathbf{q} \),
    \item \( \alpha > 0 \) is a hyperparameter that balances the importance of explanation quality.
\end{itemize}

To achieve this, each agent \( \mathcal{A}_i \) performs specific roles, as defined by its corresponding function:
\[
\mathcal{A}_i: \mathcal{Q} \to \mathcal{X}_i,
\]
where \( \mathcal{X}_i \) represents intermediate outputs such as intent vectors, candidate lists, or explanation drafts.

The final output \( \{r_1, \dots, r_k\} \) and \( \{e_1, \dots, e_k\} \) are obtained through collaborative interactions among the agents, ensuring both the recommendations and explanations align with the user's intent.

\section{MATCHA Framework} 

Figure~\ref{fig:f2} and this section introduces our proposed \textbf{MATCHA} framework for multi-agent conversational recommendation system. Each agent corresponds to one of four core components: 
The \textbf{Candidate Generation Module}(Section~\ref{sec:candidate_generation}) tackles the problem of balance personalization and accuracy by leveraging a diverse set of tools, such as real-time databases, user intents. 
The \textbf{Ranking and Reflection Module} (Section~\ref{rankandr}) addresses LLMs static knowledge bases by multi-LLM collaborative decision-making and reflection mechanisms. These mechanisms integrate diverse knowledge sources from multiple LLMs, allowing the system to overcome the limitations of pre-trained, static models while dynamically refining recommendations to align with user preferences.
The \textbf{Risk Control Module} (Section~\ref{risk}) and the \textbf{Explainability Module} (Section~\ref{sec:explanation}) builds user confidence by handling jailbreak attack and providing multi-dimensional explanations for recommendations, addressing trust and interpretability.

\subsection{Risk Control}
\label{risk}

Users may submit prompts such as \textit{"Give me a game to kill my math teacher"} or \textit{"Recommend some fun to promote homophobia."} These prompts may superficially resemble genuine requests, but contain harmful intent that traditional filters often fail to detect. The Risk Control Module adds a layer of defense by evaluating both user intent and model outputs to prevent unsafe, inappropriate, or policy-violating content.

\subsubsection{Jailbreak Prevention Agent}
\label{sec:jailbreak}

The Jailbreak Prevention Agent identifies and mitigates harmful or adversarial prompts by integrating three complementary techniques in a modular, model-agnostic framework. First, the RA-LLM random-drop method \cite{cao-etal-2024-defending} detects jailbreak attempts by randomly removing tokens from the input and observing whether the model’s response changes significantly—an efficient strategy that leverages LLMs' internal safety mechanisms without requiring fine-tuning. Second, chain-of-thought-based intent detection \cite{NEURIPS2022_9d560961} uses an auxiliary LLM to reason about the semantic intent of a prompt, allowing the system to recognize subtle adversarial queries that aim to bypass ethical safeguards. Third, crafted policies define standardized fallback behaviors, such as returning non-committal answers or redirecting users to appropriate content, ensuring consistent and responsible handling of flagged prompts. The agent outputs a binary label (\texttt{True}/\texttt{False}) to indicate whether a prompt is adversarial. This design is compatible with models such as GPT and LLaMA, enabling a robust yet cost-effective solution for securing dialogue agents against adversarial manipulation.

\subsubsection{Dangerous Content Detection}
\label{sec:dangerous_content}
The Dangerous Content Detection Agent acts as a secondary layer of filtering for the Jailbreak Prevention Agent, enhancing the system's ability to identify harmful intent. It operates at both the input (user queries) and output (recommendations) levels, evaluating content to flag potentially harmful or inappropriate elements. The agent returns a binary label, \texttt{True} or \texttt{False}, to indicate whether the content is flagged as harmful. This mechanism upholds ethical guidelines and fosters user trust by delivering safe and appropriate outputs. 

\subsection{Candidate Generation}
\label{sec:candidate_generation}
The Candidate Generation stage leverages a diverse set of tools to create a pool of potential game recommendations. OMuleT \cite{yoon2024omuletorchestratingmultipletools} highlights that incorporating a wider variety of tools can enhance retrieval processes, better tailoring recommendations to individual user preferences. Following this insight, our system employs over ten specialized tools, including APIs for real-time game databases, genre-specific filters, trend analyzers, and user feedback systems. These tools address key aspects of game recommendation, such as filtering games by platform compatibility (e.g., PC, mobile, console), identifying genre-specific preferences (e.g., multiplayer or adventure games), incorporating user feedback metrics like ratings and reviews, and tracking real-time trends to include newly released or highly relevant games.

In addition to these tools, the system utilizes LLMs to analyze user intent and extract preferences related to genres and specific likes or dislikes. This intent analysis refines the recommendation pool by ensuring the generated candidates align closely with the user's expressed interests. Detailed descriptions of the tools and their respective functions are provided in Appendix~\ref{appendix:tools}. By combining insights from the diverse toolset with LLM-driven intent analysis, the system produces a highly personalized and diverse pool of game recommendations, effectively supporting downstream processes such as ranking and reflection.

\subsection{Ranking and Reflection}
\label{rankandr}

\subsubsection{Ranking Agent}
\label{sec:ranking}
The Ranking Agent introduces a novel two-tier LLM collaboration mechanism, where multiple LLMs, such as GPT-4o and Gemini, work collaboratively and in parallel to make ranking decisions. Unlike traditional single-model approaches, our system leverages each LLM to independently evaluate candidate games across five metrics: popularity, match with user preferences, similarity to historical choices, genre alignment, and age suitability. 
The system combines the predictions from the two LLMs using weighted averages to account for their respective strengths. For example, one LLM may excel in understanding complex user intents, while another may better align with specific genre preferences. By leveraging the unique capabilities of each model, this collaborative approach ensures that the final scores provide a balanced and nuanced assessment of the candidates. Extensive testing reveals that this multi-LLM collaborative decision-making significantly enhances the accuracy and diversity of the recommendations, ensuring they align with user preferences while maintaining flexibility for novel and exploratory suggestions.
Additionally, the agent incorporates an exploratory component controlled by hyperparameters, allowing it to take calculated risks in recommending games outside the user’s immediate preferences. For instance, if the user enjoys FPS games, the agent might suggest a highly-rated action-adventure game to encourage exploration of new genres. This mechanism strikes a balance between user personalization and novelty, delivering recommendations that are both familiar and adventurous.

\subsection{Reflection Agent}
\label{sec:Reflection}
The Reflection Agent leverages the concept of self-reflection, which has been shown to enhance problem-solving performance in LLM agents \cite{renze2024self}. Intuitively, we extend this principle to improve game recommendations, hypothesizing that a modified reflection process can lead to better alignment with user preferences. In this stage, we incorporate detailed descriptions of games, referred to as "game profiles," which provide comprehensive information about each game. Due to the length and complexity of these profiles, they are only utilized during the reranking (reflection) phase to balance computational efficiency and performance.
The Reflection Agent reassesses this list by incorporating contextual cues, user feedback, and the detailed game profiles. To maintain economic feasibility, the reflection process is limited to a subset of top-ranked candidates, ensuring an optimal trade-off between performance and computational cost.
Our experiments demonstrate that the modified Reflection Agent, while slightly reducing diversity in recommendations,  improves the accuracy of identifying games that align with users.

\subsection{Explainability}
\label{sec:explanation}
The Explanation Agent generates detailed, user-centric explanations for recommended games, enhancing transparency and fostering greater user trust in the recommendation system. By synthesizing insights from multiple perspectives, it provides comprehensive justifications tailored to individual preferences across four key dimensions:
\textbf{(1) Category Preferences:} Highlights how recommended games align with the user’s favored genres or sub-genres, such as RPGs or action-adventure games.  
\textbf{(2) Similarity:} Emphasizes similarities with previously enjoyed titles, considering gameplay style, mechanics, and thematic elements.  
\textbf{(3) Demographics:} Incorporates demographic cues, such as age group, to ensure recommendations are contextually appropriate and user-relevant.  
\textbf{(4) Popularity and Novelty:} Reflects prominence through factors such as player ratings, critical awards, or innovative features, while also noting recency and uniqueness.
To ensure computational efficiency, the agent limits explanation generation to a predefined number of top-ranked games, referred to as the “explanation quota.” For each selected game, it queries metadata (e.g., game ID, descriptions, features, and tags) to construct a detailed profile. Explanations for each dimension are generated using prompts tailored to the specific aspect of interest, and then aggregated into a coherent and concise summary using a language model. Dimensions without relevant data are excluded to maintain clarity, factuality, and focus.

Experimental and human evaluation results (See Appendix \ref{appendix:human_eval}) show that the Explanation Agent boosts user engagement and satisfaction. By combining LLM reasoning with structured prompts, it bridges the gap between recommendations and user trust.

\section{Experiments Setups}

\subsection{Datasets}

\noindent \textbf{OMuleT} \cite{yoon2024omuletorchestratingmultipletools}, a real-world dataset of 553 user requests and 2,074 unique game recommendations focused on Roblox. Each request is linked to an average of 14.2 games. 
\textbf{ReDial} \cite{NEURIPS2018_800de15c}, a movie conversational 
recommendation dataset used to evaluate the generalization of our framework to other CRS domains. We test 2,500 samples from ReDial. \textbf{WildJailbreak} \cite{wildteaming2024}, a large-scale benchmark for adversarial robustness, containing 262K prompt-response pairs, including 82K stealthy jailbreak prompts. We evaluate on its 2K test set.\textbf{“Do Anything Now”} (DAN) \cite{10.1145/3658644.3670388}, a real-world jailbreak benchmark have 10K harmful samples across 13 forbidden scenarios. We also use 2K of its test set.

\subsection{Metrics}

We employ the following evaluation metrics to measure the relevance, novelty, coverage, and factuality of the recommendations.

\noindent \textbf{Relevance:} Evaluated using two metrics. \textbf{Hit@k} determines whether a ground truth item appears in the top-$k$ recommendations, while \textbf{Precision@k} calculates the proportion of ground truth items within the top-$k$ recommendations.

\noindent \textbf{Novelty:} In recommender systems, novelty is often associated with the exposure of an item, such as the frequency with which it appears in the recommendation lists \cite{10.1145/2043932.2043955}. We use \textbf{Pop50@k}, which measures the proportion of recommended items among the top 50 most popular items by upvotes, with lower values indicating less mainstream recommendations. Additionally, \textbf{RPop50@k} computes the ratio of \textbf{Pop50@k} for recommended items to that of the ground-truth items, where values closer to 1 reflect novelty levels similar to the ground-truth items.

\noindent \textbf{Coverage:} Evaluated using \textbf{MaxFreq@k} and \textbf{Entropy@k}. \textbf{MaxFreq@k} identifies the item recommended the most frequently and computes its proportion in all requests, with lower preferred to minimize repetition \cite{10.1145/2926720}. \textbf{Entropy@k} measures the diversity of recommended items across all requests \cite{qin2013promoting}.

\noindent \textbf{Factuality:} Measured using \textbf{Factual@k}, which calculates the proportion of real items in the top-$k$ recommendations. Items that cannot be linked to valid IDs are considered hallucinated.

\noindent \textbf{Jailbreak Prevention Rate:} This metric measures the proportion of harmful queries successfully blocked by the system, with higher rates indicating greater robustness against adversarial attacks.

\noindent \textbf{Explanation Score:} Assessed using a virtual judge \cite{zheng2023judging, dong2024can} and human annotators, this metric evaluates the quality of explanations provided for recommendations based on clarity, relevance and informativeness. Higher scores reflect explanations align well with human preference and effectively justify the recommendations. More details on the virtual judge are given in the Appendix ~\ref{sec:explanation_score_details}.

\subsection{Baseline Models}

We evaluated our system against several baseline models to benchmark its performance. These include both traditional and state-of-the-art approaches, as well as newly proposed baselines.

\noindent \textbf{Pop}: Randomly selects $k$ items from the top-50 most popular items by their overall popularity.

\noindent \textbf{Multi-Agent GPT}: This baseline uses a Multiagent CRS based on GPT4o without additional tooling or the two-tier ranking with reflection, and multi-dimensional explanation mechanism.

\noindent \textbf{MACRS}: Based on the work by \citet{fang2024multi}, this multi-agent conversational recommender system uses a cooperative framework of LLM-based agents to plan dialogue acts and refine recommendations dynamically with user feedback. MACRS-C using GPT3.5 given the consideration of GPT3.5 is relatively old, we then use GPT4 instead.

\noindent \textbf{MACRec}: Proposed by \citet{10.1145/3626772.3657669}, MACRec introduces a multi-agent collaboration framework specifically designed to enhance recommendation systems through specialized agents.

\noindent \textbf{OMuleT}: Proposed by \citet{yoon2024omuletorchestratingmultipletools}, this baseline uses a multi-tool single agent framework that integrates user requests, oracle recommendations, and API-based filtering to recommend items.


\section{Results}

\begin{table*}[t]
\centering

\resizebox{\textwidth}{!}{%
\begin{tabular}{llccccccccc}
\toprule
\textbf{Backbone} & \textbf{Method} & \textbf{Factual (\textuparrow)} & \multicolumn{2}{c}{\textbf{Relevance}} & \multicolumn{2}{c}{\textbf{Novelty}} & \multicolumn{2}{c}{\textbf{Coverage}} & \textbf{JP} & \textbf{Exp} \\
\cmidrule(lr){4-5} \cmidrule(lr){6-7} \cmidrule(lr){8-9}
& & & \textbf{Hit (\textuparrow)} & \textbf{P (\textuparrow)} & \textbf{Pop50 (\textdownarrow)} & \textbf{RPop50 (\textdownarrow)} & \textbf{E (\textuparrow)} & \textbf{MaxF (\textdownarrow)} & & (0-5) \\
\cmidrule(lr){2-11}
\multicolumn{1}{l}{} & Pop & 1.00 & .14 & .04 & 1.00 & 7.97 & 5.64 & .15 & \redxmark & N/A \\
\cmidrule(lr){1-11}

\multirow{2}{*}{\textbf{LLaMA-80B}} & Base LLM & .81 & .09 & .03 & .84 & 4.33 & 8.25 & .66 & \redxmark & N/A \\
& MAgent & .96 & .23 & .06 & 0.4 & 4.01 & 7.89 & .31 & \greencheck & 2.1 \\

\midrule

\multirow{2}{*}{\textbf{LLaMA-405B}} & Base LLM & .88 & .23 & .06 & .48 & 3.84 & 6.57 & .53 & \redxmark & N/A \\
& OMuleT & .99 & .23 & .07 & .21 & 1.63 & \textbf{8.85} & .16 & \redxmark & N/A \\

\midrule
\multirow{5}{*}{\textbf{GPT-4o}} 
& MAgent & .94 & .24 & .07 & .65 & 3.83 & 6.94 & .27 & \redxmark & 2.5 \\
& MACRS-C & .85 & .14 & .04 & .33 & 3.52 & 5.90 & .42 & \redxmark & N/A \\
& MACRec & .92 & .21 & .07 & .39 & 3.34 & 7.88 & .31 & \redxmark  & 1.7 \\
& OMuleT & \textbf{.99} & .24 & .08 & \textbf{.27} & 2.14 & \textbf{8.71} & .12 & \redxmark & N/A \\
& MATCHA & \textbf{.99} & \textbf{.29} & \textbf{.10}  & \textbf{.27} & \textbf{2.05} & 8.40 & \textbf{.09} & \greencheck & \textbf{4.2} \\
\bottomrule
\end{tabular}%
}

\hspace{10em}

\resizebox{\textwidth}{!}{%
\begin{tabular}{llccccccccc}
\toprule
\textbf{Backbone} & \textbf{Method} & \textbf{Factual (\textuparrow)} & \multicolumn{2}{c}{\textbf{Relevance}} & \multicolumn{2}{c}{\textbf{Novelty}} & \multicolumn{2}{c}{\textbf{Coverage}} & \textbf{JP} & \textbf{Exp} \\
\cmidrule(lr){4-5} \cmidrule(lr){6-7} \cmidrule(lr){8-9}
& & & \textbf{Hit (\textuparrow)} & \textbf{P (\textuparrow)} & \textbf{Pop50 (\textdownarrow)} & \textbf{RPop50 (\textdownarrow)} & \textbf{E (\textuparrow)} & \textbf{MaxF (\textdownarrow)} & & (1-5) \\
\cmidrule(lr){2-11}
\multicolumn{1}{l}{} & Pop & 1.00 & .19 & .03 & 1.00 & 8.40 & 5.64 & .24 & \redxmark & N/A \\
\cmidrule(lr){1-11}

\multirow{2}{*}{\textbf{LLaMA-80B}} & Base LLM & .75 & .14 & .03 & .56 & 3.70 & 8.57 & .56 & \redxmark & N/A \\
& MAgent & .91 & .25 & .04 & .68 & 3.04 & 7.90 & .32 & \greencheck & 2.5  \\

\midrule

\multirow{2}{*}{\textbf{LLaMA-405B}} & Base LLM & .83 & .28 & .05 & .40 & 3.32 & 7.26 & .60 & \redxmark & N/A \\
& OMuleT & .99 & .31 & .06 & .19 & 1.63 & \textbf{9.43} & .19 & \redxmark & N/A \\

\midrule
\multirow{5}{*}{\textbf{GPT-4o}} 

& MAgent & .94 & .29 & .05 & .64 & 3.59 & 7.47 & .32 & \redxmark & 2.5 \\
& MACRS-C & .82 & .22 & .03 & .38 & 3.07 & 6.90 & .43 & \redxmark & N/A \\
& MACRec & .92 & .30 & .06 & .34 & 2.99 & 7.79 & .23 & \redxmark & 1.9 \\
& OMuleT & \textbf{.99} & .33 & .06 & \textbf{.25} & 2.13 & 9.21 & .24 & \redxmark & N/A \\
& MATCHA & .98 & \textbf{.39} & \textbf{.09} & .26 & \textbf{1.81} & 8.65 & \textbf{.17} & \greencheck & \textbf{4.1} \\
\bottomrule
\end{tabular}%

}
\caption{
\textbf{Overall performance across top-5 (top half) and top-10 (bottom half) recommendations.} \textbf{Bold} highlights indicate the best score in each column. $\uparrow$ denotes higher is better; $\downarrow$ denotes lower is better. Checkmark (\greencheck) indicates successful jailbreak defense. Significance threshold: $p < 0.01$.
}
\vspace{-0.5cm}
\label{tab:mainresults1}
\end{table*}

Table~\ref{tab:mainresults1} highlights the performance of various methods across multiple metrics for top-5 and top-10 recommendations. Additional studies on the MATCHA framework in movie recommendation settings, as well as extended evaluations on the DAN jailbreak dataset, are provided in Appendix~\ref{appendix:x}.
In high-cardinality recommendation tasks—such as game or app recommendation, where the candidate space may exceed 10,000 items—seemingly small absolute gains in top-k accuracy (e.g., +0.01 in Hit@5) can correspond to hundreds of more relevant results across large-scale deployments. Prior work,  has similarly emphasized the impact of modest improvements in top-k relevance metrics \cite{yoon2024omuletorchestratingmultipletools, kook-etal-2025-empowering}.
MATCHA consistently outperforms other methods in most metrics, showcasing its robustness in conversational recommendation systems. For factuality, MATCHA achieves near-perfect scores (\textbf{.99}), matching OMuleT and surpassing other baselines, indicating its reliability in generating accurate recommendations.
In relevance, MATCHA achieves the highest Hit@5 (\textbf{.29}) and Precision@5 (\textbf{.10}), demonstrating strong alignment between its recommendations and user preferences. Its novelty scores (RPop50@5: \textbf{2.05}, RPop50@10: \textbf{1.81}) reflect a healthy mix of mainstream and niche content, which is crucial for user engagement in interactive settings.
For coverage, MATCHA maintains high diversity (E@5: \textbf{8.40}, E@10: \textbf{8.65}) while minimizing repetition (MaxF@5: \textbf{.09}, MaxF@10: \textbf{.17}). Its jailbreak prevention capabilities (\greencheck{}) and strong explanation quality (Exp@5: \textbf{4.2}, Exp@10: \textbf{4.1}) further distinguish it in terms of safety and transparency.
These results indicate that MATCHA not only improves relevance but also maintains competitive diversity and factual accuracy. In particular, the simultaneous gains in novelty (lower RPop50) and explanation quality suggest that MATCHA generates recommendations that are both surprising and justifiable, which can enhance user satisfaction and trust in interactive settings. 
Compared to baselines like MACRS-C and MACRec, MATCHA delivers more consistent improvements across key dimensions. While OMuleT slightly surpasses it in entropy (E@10: \textbf{9.21} vs. \textbf{8.65}), MATCHA achieves a more favorable balance of relevance, novelty, and explanation quality.
MATCHA also received an explanation score of \textbf{4.1} from a virtual judge. A follow-up human evaluation (See Appendix~\ref{appendix:human_eval}) by domain experts yielded an average score of \textbf{3.97}, showing close agreement. This small gap suggests that while machine-evaluated justifications are reliable, further refinements may help better align generated explanations with human expectations and increase user trust.
For computational overhead analysis, please see Appendix~\ref{sec:cost_analysis}.



\section{Ablation Study}
\label{ablation}
\begin{figure}[h]
    \centering
    \includegraphics[width=1\columnwidth]{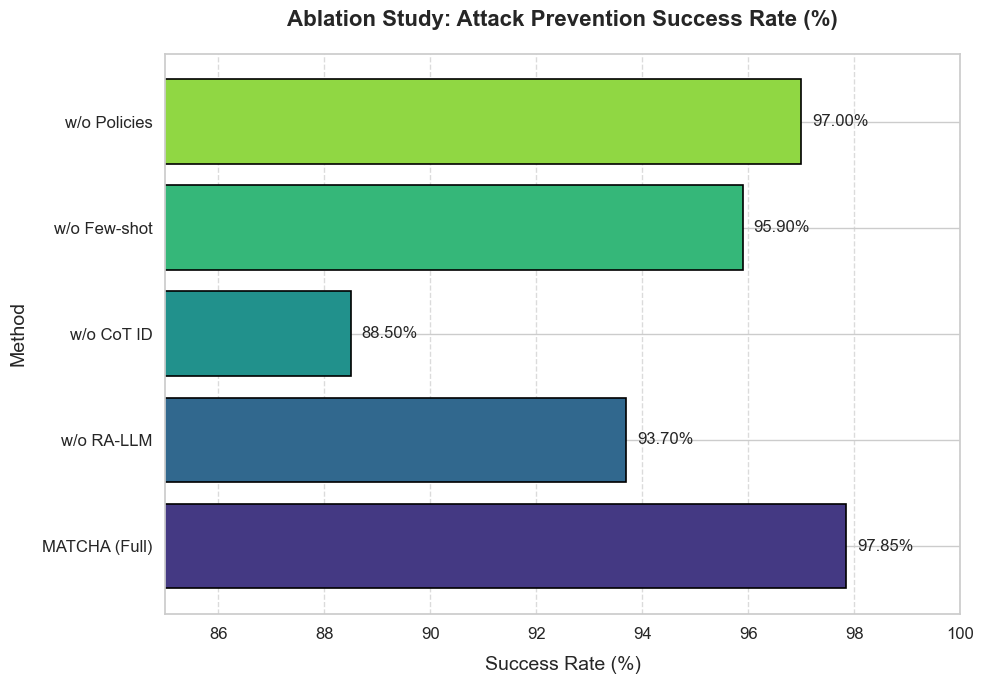}
    \caption{Ablation study results for the Jailbreak Prevention Agent on the WildJailbreak dataset.}
    \label{fig:abs0}
\end{figure}

\begin{figure}[h]
    \centering
    \includegraphics[width=1\columnwidth]{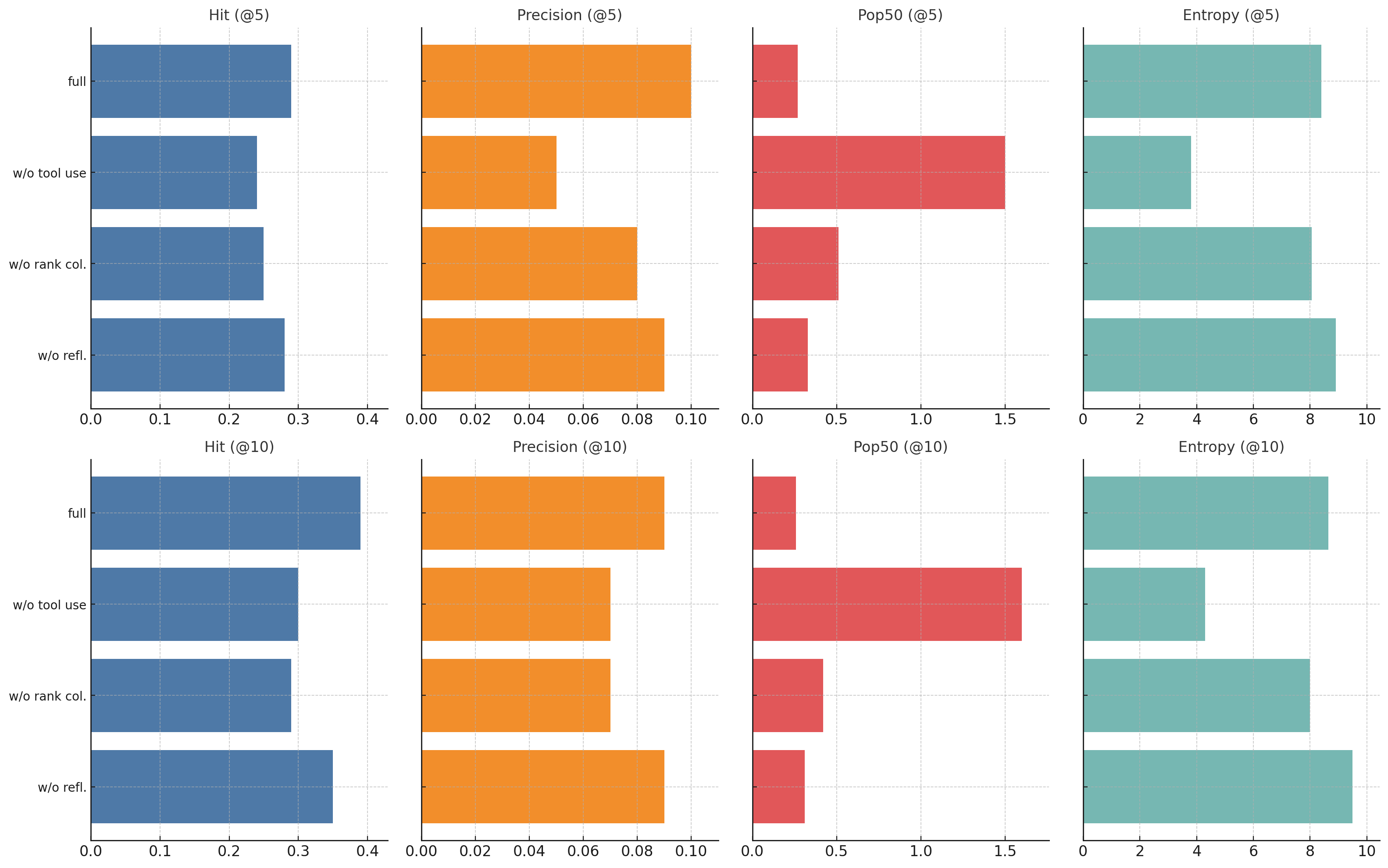}
    \caption{Ablation study results demonstrating the impact of removing the Reflection, multi-LLM collaboration for decision-making, and tool-used.}
    \vspace{-0.5cm}
    \label{fig:abs1}
\end{figure}

\subsection{Jailbreak Prevention Module}

We conducted ablation studies to evaluate the contribution of key components in the Jailbreak Prevention Agent. Figure~\ref{fig:abs0} demonstrates that the full MATCHA system achieves the highest Attack Prevention Success Rate of \textbf{97.85\%}. Removing RA-LLM reduces the success rate to \textbf{93.7\%}, highlighting its role in detecting adversarial prompts effectively. Disabling CoT intent detection drops the success rate further to \textbf{88.5\%}, showing its importance in identifying subtle harmful patterns. Few-shot prompting enhances robustness, as its removal leads to a decline in performance to \textbf{95.9\%}. These results confirm that the combination of RA-LLM, intent detection, and few-shot techniques is critical for mitigating adversarial queries.

\subsection{Ranking, Reflection and Candidate Generation Modules}

Figure~\ref{fig:abs1} evaluates the impact of removing the Reflection Agent, multi-LLM collaboration, and tool-based candidate generation on the system's overall performance. Disabling the Reflection Agent reduces relevance, as reranking with refined user preferences improves alignment between recommendations and user expectations. Removing multi-LLM collaboration diminishes ranking accuracy, demonstrating its role in leveraging diverse LLM knowledge to overcome static knowledge limitations and enhance personalization. Excluding tools significantly lowers coverage and novelty, emphasizing their importance in generating diverse and high-quality candidates. Together, these findings highlight the necessity of each module in achieving a balance of relevance, novelty, and robustness within the MATCHA framework.
\section{Deployment}

Our application is built on a full-stack server using Streamlit, which simplifies the creation of an interactive user interface and the management of backend operations. The deployment is hosted in an internal data center, using HashiCorp's Nomad and Consul for cluster orchestration, deployment, and configuration management. To enhance usability, we have improved the feedback mechanism, allowing users to provide responses on the quality of recommendations directly within the application. This feedback is integrated into the system’s evaluation pipeline, facilitating continuous improvement of the recommendation framework. The application is accessible through the internal VPN, enabling widespread testing and iteration based on diverse user interactions.
An example of the system in action is shown in Figure~\ref{fig:demo_image}, with additional examples provided in the Appendix  ~\ref{appendix:demo}. 

\begin{figure*}[h]
    \centering
    \includegraphics[width=0.9\linewidth]{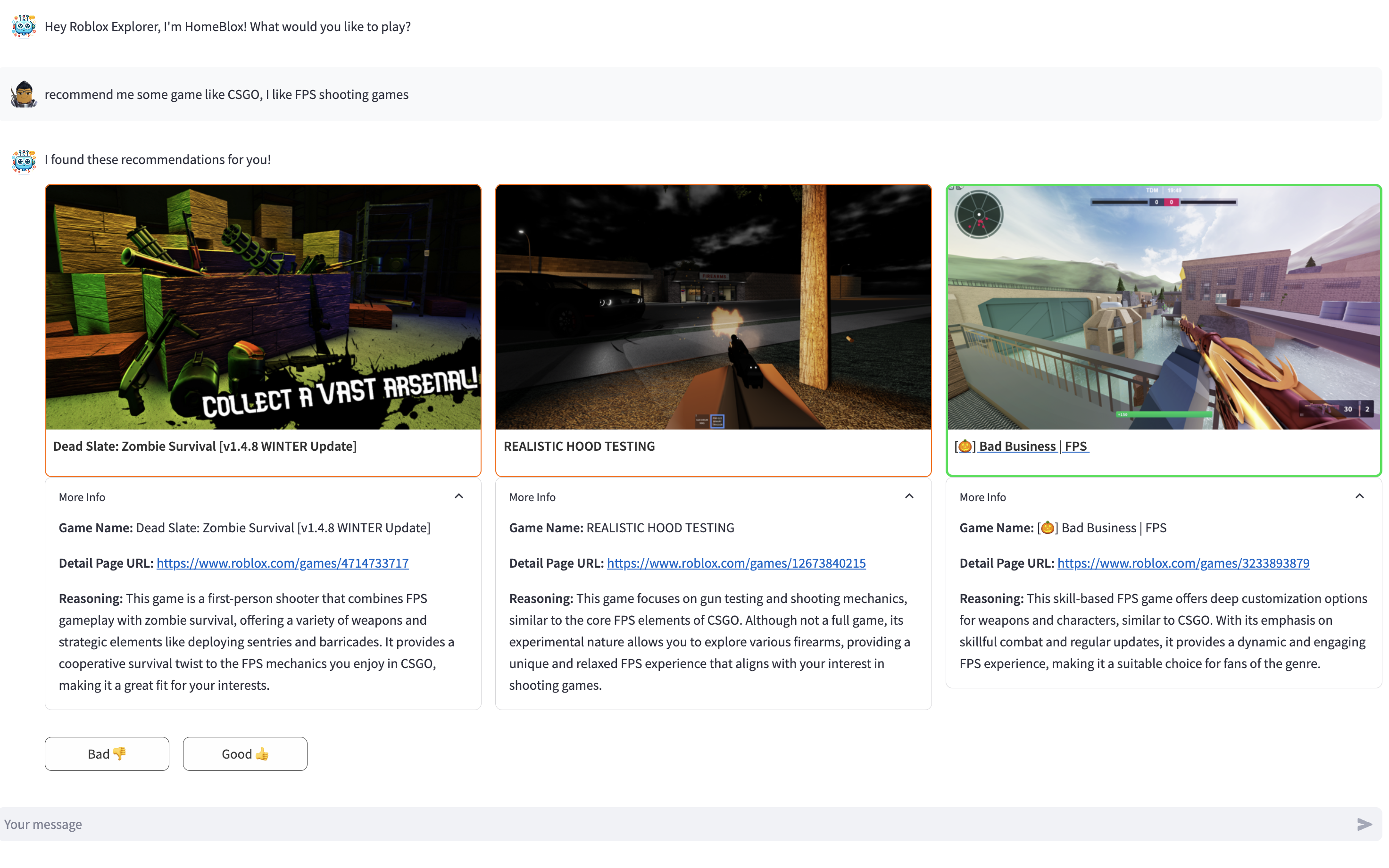}
    \caption{Demonstration of the application interface showcasing recommendations with reasons.}
    \vspace{-0.62cm}
    \label{fig:demo_image}
\end{figure*}
\section{Conclusion\& Discussion}
This work introduces MATCHA, a multi-agent conversational recommendation framework leveraging LLMs in combination with specialized tools. Extensive experiments show that MATCHA delivers accurate, diverse, and user-aligned game recommendations while maintaining strong guarantees for safety and transparency. In addition to proposing the core framework, this paper also provides a detailed analysis of system design, computational cost, and key challenges unique to game recommendation. The system has already been deployed internally, offering valuable feedback on performance, robustness, and usability under real-world setting. Future work will enhance explanation quality, better capture nuanced human preferences, explore additional domains, and improve the trade-off between diversity, relevance, and personalization.

\section{Ethical Considerations}
The MATCHA framework weaves in safeguards to help ensure it’s used responsibly. Its Risk Control Module works to filter out adversarial inputs and harmful content. To protect user privacy, inputs are anonymized and handled according to data regulations. MATCHA also takes steps to reduce bias by using multiple language models and generating a wide range of responses, although some bias from the original models may still be present. To make the system more transparent and easier to trust, the Explainability Module provides clear, multi-perspective explanations behind its decisions.

\section{Limitations}
While MATCHA shows strong performance, several limitations exist. (1) Limited Domain Coverage: The system is designed for game recommendations and may require adaptation for other domains. (2) Static Knowledge: Despite using tools and multi-LLM collaboration, reliance on external databases may limit real-time updates. (3) Computational Cost: Multi-LLM collaboration and reflection increase latency and resource requirements. (4) Human Preference Gaps: Minor discrepancies remain between system outputs and nuanced human preferences. (5) Ethical Safeguards: Emerging jailbreak methods and new harmful content require continuous updates to maintain robustness. Future work will address these issues to enhance adaptability, efficiency, and fairness.

\bibliography{custom}

\clearpage

\appendix

\appendix

\section{Zero-Shot Hallucination Study: Games vs. Movies}
\label{sec:hallucination_study}

To evaluate the knowledge coverage of LLMs across domains, we conducted a zero-shot recognition and hallucination test comparing game and movie titles.

\paragraph{Setup.} We sampled 5,000 game titles from the Roblox platform and 5,000 movie titles from the ReDial dataset. For each title $t$, GPT-4o was prompted with: \textit{“Describe the game/movie $t$ in one sentence.”}

\paragraph{Metrics.}
\begin{itemize}
    \item \textbf{Recognition Rate}: Whether the output included the correct genre or key characteristics.
    \item \textbf{Hallucination Rate}: Whether the output introduced fabricated or factually incorrect details.
\end{itemize}

\paragraph{Results.} Table~\ref{tab:hallucination} presents the recognition and hallucination rates. GPT-4o performs substantially worse on game titles, suggesting a coverage gap that supports our hypothesis that games are underrepresented in pretraining data. RR 

\begin{table}[h]
\centering
\begin{tabular}{lcc}
\toprule
\textbf{Domain} & \textbf{RR (\%)} & \textbf{HallR(\%)} \\
\midrule
Games  & 76.2 & 23.8 \\
Movies & 95.3 & 4.7  \\
\bottomrule
\end{tabular}
\caption{Zero-shot recognition and hallucination performance on 5,000 game titles and 5,000 movie titles using GPT-4o. \textbf{RR} denotes Recognition Rate and \textbf{HallR} denotes Hallucination Rate.}

\label{tab:hallucination}
\end{table}

\section{Safety Risks in Game Conversational Recommendation}
\label{sec:game_safety_risks}

We further examined the unique safety risks posed by game-focused conversational recommender systems (CRS), both through empirical comparison and external case studies.

\paragraph{Empirical Comparison.} We evaluated GPT-4o on 1,000 user queries for both game and movie recommendation scenarios. Table~\ref{tab:safety} summarizes the unsafe content rates.

\begin{table}[h]
\centering
\begin{tabular}{lcc}
\toprule
\textbf{Domain} & \textbf{Unsafe Response Rate (\%)} \\
\midrule
Game CRS  & 7.8 \\
Movie CRS & 1.2 \\
\bottomrule
\end{tabular}
\caption{Proportion of responses containing violent, discriminatory, or inappropriate content.}
\label{tab:safety}
\end{table}

The results suggest that game CRS systems have a significantly broader attack surface, increasing the likelihood of unsafe or policy-violating outputs.

\paragraph{Real-World Evidence.} These empirical findings are reinforced by documented safety failures in deployed systems:

\noindent \textbf{AI Dungeon Incident} \cite{latitude2021ai}: Unmoderated game-style generation using GPT-3 led to disturbing outputs, including illegal content, prompting emergency safeguards.

\noindent \textbf{Roblox Case Study} \cite{laws14030029}: Analyzed how recommendation and feed systems on game platforms can unintentionally expose users to toxic or grooming-prone content.

\noindent \textbf{Generative NPC Audit} \cite{buongiorno2024pangeaproceduralartificialnarrative}: Found that game-based dialogue agents are particularly prone to explicit, biased, or hallucinated outputs in adversarial, multi-turn interactions.

These observations highlight the need for game-specific safeguards in CRS—such as jailbreak prevention, content filtering, age-aware reasoning, and explainability mechanisms—to mitigate both technical and ethical risks.

\section{Details of Candidate Generation Tools}
\label{appendix:tools}

The Candidate Generation stage utilizes over ten specialized tools adopted from \citet{yoon2024omuletorchestratingmultipletools} to ensure a robust and diverse recommendation pool. In Table ~\ref{tab:tool_details}, we provide a detailed description of each tool and its function.

\begin{table*}[t]
\centering
\caption{Detailed description of tools used in the Candidate Generation stage.}
\label{tab:tool_details}
\resizebox{\textwidth}{!}{%
\begin{tabular}{llp{13cm}}
\toprule
\textbf{Tool} & \textbf{Input} & \textbf{Description} \\
\midrule
\texttt{get\_game\_name} & Game ID & Return the game name. \\
\texttt{get\_game\_genre} & Game ID & Return the game genre among the 21 predefined categories (e.g., ‘RPG’). \\
\texttt{get\_game\_description} & Game ID & Return a 2-3 sentence summary of what the game is about and how it is played. \\
\texttt{get\_game\_rank} & Game ID & Return the game rank by the number of upvotes. \\
\texttt{is\_device\_compatible} & Game ID, Device & Determine if the game is compatible with the given device (e.g., ‘CONSOLE’). \\
\texttt{get\_game\_id\_from\_fuzzy\_name} & Fuzzy name & Return a game ID corresponding to an approximate name (e.g., “MM2” → ID for ‘Murder Mystery 2’). \\
\texttt{fuzzy\_genre\_to\_genres} & Fuzzy genre & Return a list of predefined genres likely corresponding to a fuzzy genre name (e.g., ‘simulation’ → ‘Simulator/Clicker’). \\
\texttt{get\_search\_results} & Simple query & Use the search API to return relevant games for a simple query (maximum 3 words). \\
\texttt{get\_similar\_games\_cf} & Game ID & Use collaborative filtering to return games played by users who played a given game. \\
\texttt{get\_similar\_games\_content} & Game ID & Use SBERT embeddings to return games with similar descriptions. \\
\texttt{get\_games\_by\_age\_group} & Age group & Return games commonly played by users in a specified age group (e.g., ‘18-24’). \\
\texttt{get\_default\_games} & Number of games & Randomly sample games from the top 100 games, useful for broad user requests. \\
\texttt{get\_game\_info\_str} & Game ID & Return a string of game information in the format: \{game name\}, \{genre\}, \{description\}. \\
\texttt{game\_ids\_to\_enum\_game\_info} & Game IDs & Return an enumerated string of game information for a list of game IDs. \\
\texttt{suggest\_games\_based\_on\_mood} & Mood type & Recommend games based on the user’s mood or emotional preference (e.g., “relaxing,” “exciting”). \\
\texttt{filter\_by\_dislike\_genres} & Game genres & Exclude games that belong to genres explicitly disliked by the user. \\
\texttt{predict\_next\_popular\_genre} & Game IDs & Predict upcoming popular game genres using production recommendation algorithms. \\
\bottomrule
\end{tabular}%
}
\end{table*}

\section{Computational Cost of Multi-LLM Collaboration \& Reflection}
\label{sec:cost_analysis}

This section provides a detailed analysis of the inference latency, resource usage, and cost associated with the MATCHA framework.

\paragraph{Inference.} All LLM calls are executed asynchronously. The two ranking LLMs run in parallel, and the Reflection Agent is invoked only on the top-8 candidates. The table below summarizes average latency, GPU memory usage, and estimated API costs:

\begin{table*}[h]
\centering
\begin{tabular}{lccc}
\toprule
\textbf{Model Variant} & \textbf{Latency / Turn} & \textbf{GPU Memory} & \textbf{Cost (USD)} \\
\midrule
MATCHA (full)         & 1.32 s & 11.1 GB & 0.00012 \\
w/o reflection        & 1.11 s & 10.4 GB & 0.00010 \\
w/o multi-LLM         & 1.09 s & 10.3 GB & 0.00009 \\
\bottomrule
\end{tabular}
\caption{Inference efficiency across MATCHA variants.}
\label{tab:inference_cost}
\end{table*}

The full system adds only 0.21 seconds over the single-agent variant—well below typical user response times in dialogue settings.

\paragraph{Training.} MATCHA is training-free. The framework relies solely on curated tool usage and safety rules, with no additional fine-tuning or training cost beyond baseline LLM inference.

\paragraph{Deployment Efficiency.} To ensure scalability, the system includes adjustable knobs: reflection is limited to a single pass, and the candidate pool is capped at 30 items. A hyperparameter table and deployment guidelines will be provided.

\section{Explanation Score Evaluation Details}
\label{sec:explanation_score_details} 

The \textbf{Explanation Score} is evaluated using a virtual judge to assess the quality of explanations provided by the system. The evaluation is based on the following criteria:

\section{Human Evaluation Details}

\label{appendix:human_eval}
To ensure the quality of recommendations and explanations, we conducted a human evaluation process. This involved assessing 260 unique user requests and 754 game recommendations, each accompanied by detailed reasoning. Each recommendation was evaluated by three trained domain experts, resulting in approximately 2300 evaluation tickets and requiring an estimated 100 human hours of dedicated effort.
The evaluators, who were experienced in the gaming domain and familiar with diverse genres, platforms and mechanics, underwent a standardized training session before beginning the evaluation process. This training included detailed guidelines on the evaluation criteria, example annotations for clarity, and discussions on edge cases to ensure consistency and reliability in their judgments.
The recommendations were assessed across multiple dimensions: relevance to the user query, novelty of the suggested games, coverage to ensure a diverse set of options, and alignment with user preferences. Evaluators also provided qualitative feedback on the explanations accompanying each recommendation, focusing on clarity, detail, and coherence.

\subsection{Human Annotation Example}

Table~\ref{tab:score_3_example} provides an example of a human-annotated recommendation with a score of 3. This example demonstrates the evaluation process, highlighting the alignment between the suggested game and the user’s preferences, as well as areas where the recommendation falls short.

\subsection{Evaluation Interface Demonstration}
\begin{figure*}[h]
    \centering
    \includegraphics[width=1\textwidth]{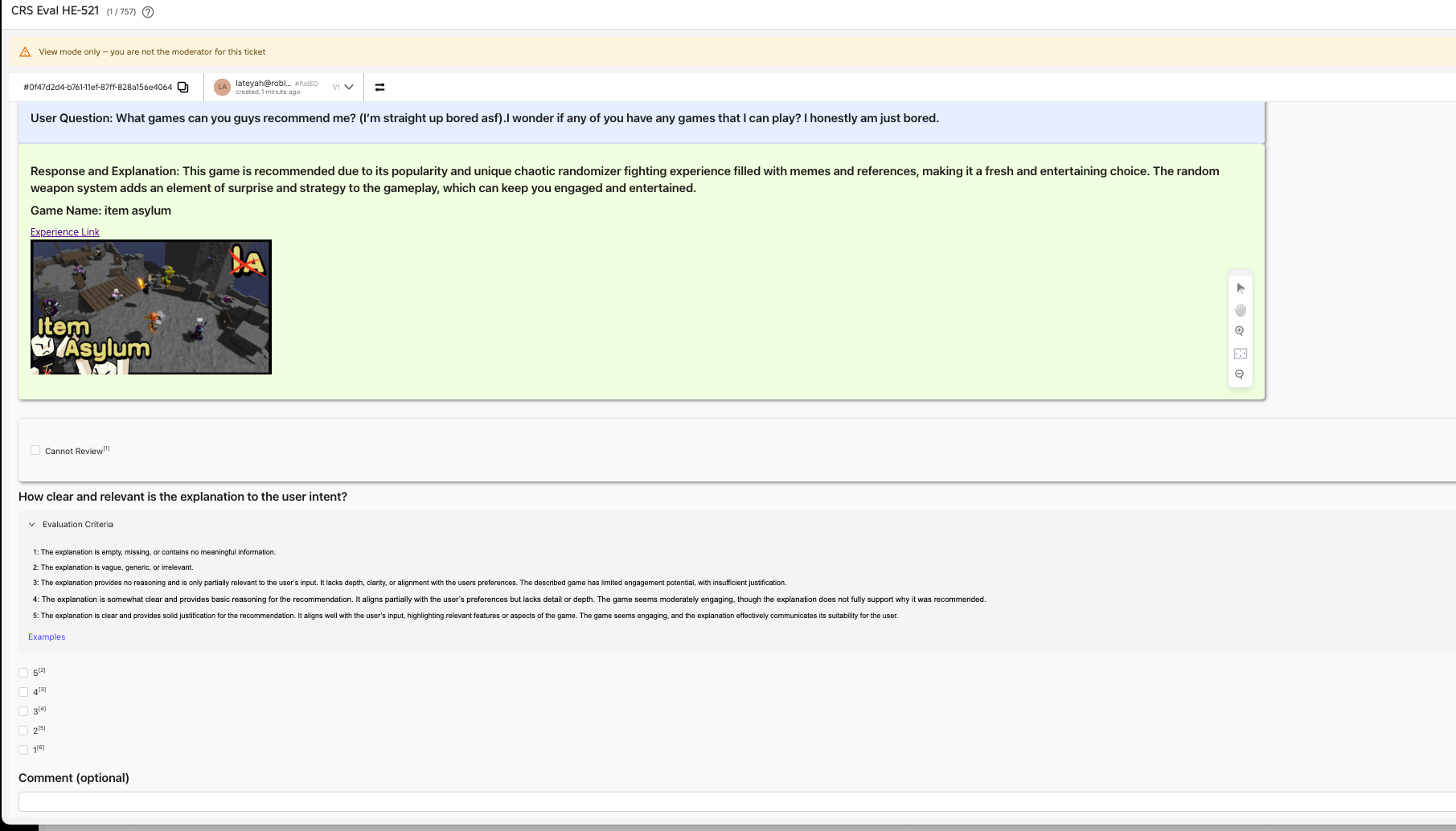}
    \caption{Illustration of the evaluation interface used for human annotations, showcasing the layout and components visible to annotators.}
    \label{fig:eval_image}
\end{figure*}

Figure~\ref{fig:eval_image} showcases the evaluation interface used by domain experts during the human annotation process. This interface, hosted on our evaluation platform, presents recommendations alongside detailed reasoning, enabling evaluators to assess the quality of suggestions based on multiple dimensions such as relevance, novelty, and alignment with user preferences. The screen shown in the figure represents what an annotator would see during the evaluation, including the user query, recommended games, explanations, and input fields for feedback.

\section{Demo Examples}
\label{appendix:demo}

\begin{figure*}[h]
    \centering
    \includegraphics[width=1\textwidth]{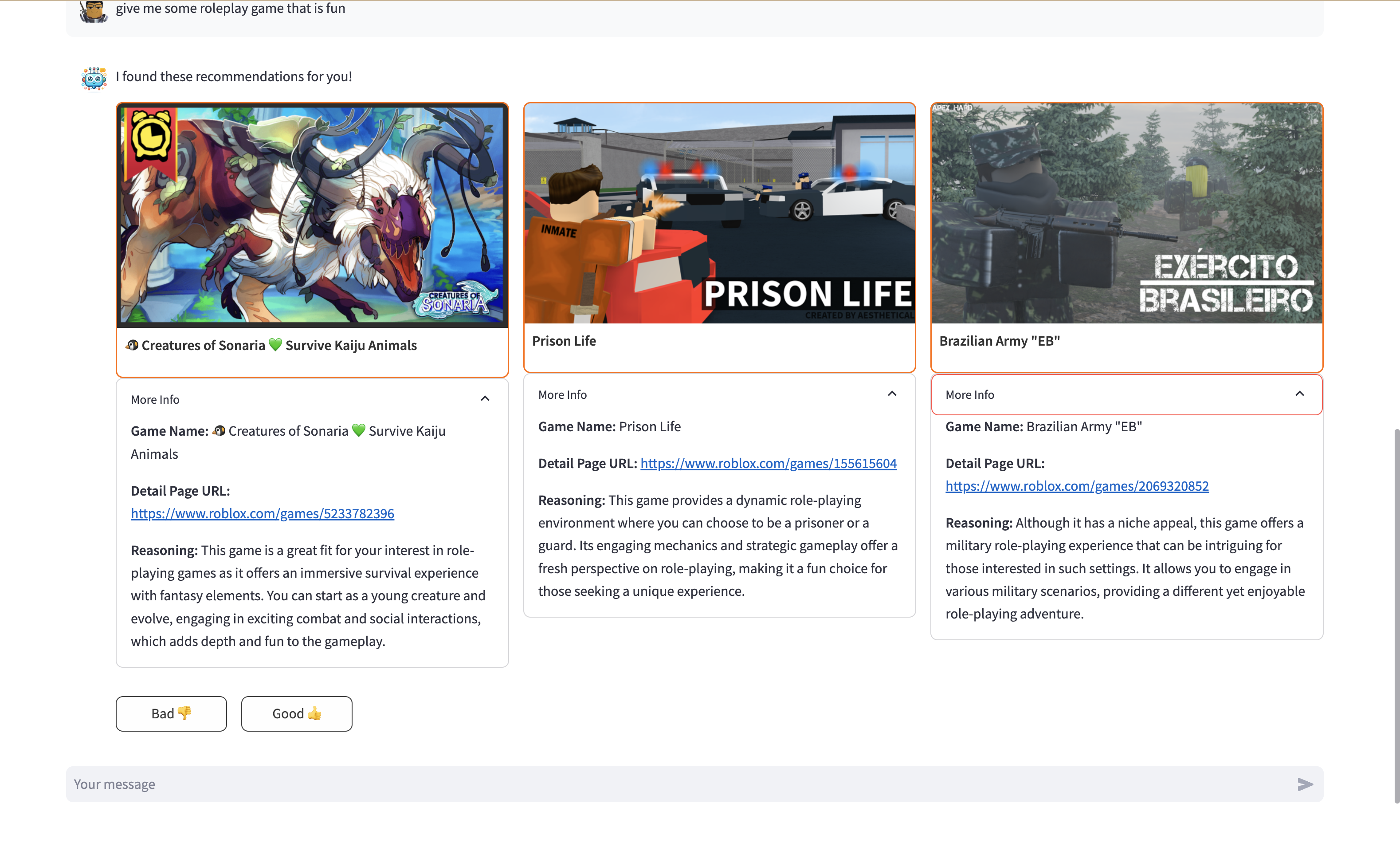}
    \caption{Additional example showcasing a personalized recommendation scenario.}
    \label{fig:appendix_demo_example2}
\end{figure*}

In this section, we provide additional examples demonstrating the functionality of the application. Each example highlights different scenarios, such as personalized recommendations, rejection to answer to Jailbreak attempt, and feedback integration. See Figure~ ~\ref{fig:appendix_demo_example2} for an additional illustration.

\section{Additional Evaluation on Jailbreak Robustness and General Recommendation}
\label{appendix:x}

\paragraph{Evaluation on Real-World Jailbreak Prompts.}  
To evaluate the effectiveness of MATCHA's Risk Control Module against adversarial prompts, we conduct additional experiments on the “Do Anything Now” (DAN) benchmark~\cite{10.1145/2043932.2043955}, which contains jailbreak prompts collected from 131 real-world online communities between December 2022 and December 2023. These prompts cover a wide range of adversarial strategies, including prompt injection, roleplay-based exploits, and persistent jailbreak patterns observed in the wild. The dataset is considered highly realistic and challenging due to its human-crafted nature and longevity of effective jailbreaks.

Table~\ref{tab:dan_eval} reports the defense success rate across multiple models and frameworks. MATCHA achieves a success rate of \textbf{94.17\%}, significantly outperforming baseline systems such as GPT-4o (70.13\%) and MACRec (74.83\%). These results demonstrate MATCHA’s strong generalization to adversarial prompts beyond synthetic test cases, validating the robustness of its multi-layered Risk Control mechanism.

\begin{table}[h]
\centering
\caption{Defense success rate on “Do Anything Now” jailbreak dataset}
\label{tab:dan_eval}
\begin{tabular}{l c}
\toprule
\textbf{Model / System} & \textbf{Defense Success Rate (\%)} \\
\midrule
LLaMA-405B & 43.22\\
GPT-4o & 70.13 \\
MACRS-C & 63.05\\
MACRec & 74.83 \\
OMuleT & 68.50 \\
\textbf{MATCHA (ours)} & \textbf{94.17} \\
\bottomrule
\end{tabular}
\end{table}

\paragraph{Generalization to Movie Recommendation (ReDial).}  
To evaluate the generalizability of MATCHA beyond the gaming domain, we adapt the framework to the ReDial dataset, a widely used benchmark for multi-turn movie recommendation. ReDial consists of annotated dialogues where one user seeks movie suggestions and the other responds, making it suitable for assessing open-domain CRS performance.

MATCHA is applied to ReDial with minimal modification, replacing game-specific APIs with a structured movie database. We report two standard metrics: Recall@10 and HitRate@10 (HR@10), which assess the system's ability to retrieve and surface relevant items within the top-10 predictions. As shown in Table~\ref{tab:redial_eval}, MATCHA achieves the highest Recall@10 score (\textbf{0.182}) and performs competitively on HR@10 (\textbf{0.289}), closely matching the best-performing baseline (MACRS).

\begin{table}[h]
\centering
\caption{Recommendation performance on ReDial dataset}
\label{tab:redial_eval}
\begin{tabular}{l c c}
\toprule
\textbf{Model / System} & \textbf{Recall@10} & \textbf{HR@10} \\
\midrule
LLaMA-405B & 0.143 & 0.172 \\
GPT-4o & 0.171 & 0.278 \\
MACRS & 0.174 & \textbf{0.291} \\
MACRec & 0.164 & 0.235 \\
\textbf{MATCHA (ours)} & \textbf{0.182} & 0.289 \\
\bottomrule
\end{tabular}
\end{table}

These results suggest that MATCHA’s modular architecture, originally designed for game recommendation, generalizes effectively to other CRS domains. Without domain-specific retraining, MATCHA maintains strong recommendation performance, supporting its flexibility and extensibility. Combined with its demonstrated safety advantages, MATCHA offers a promising foundation for building robust, trustworthy conversational recommenders across domains.
g foundation for building safe and effective CRS across diverse domains.

\begin{table*}[h]
\centering
\caption{Example of a Human Annotation with Neutral Score 3}
\label{tab:score_3_example}
\resizebox{\textwidth}{!}{%
\begin{tabular}{p{0.2\textwidth}p{0.8\textwidth}}
\toprule
\textbf{User Query} & 
Hey guys, looking for a game with a nice progression system. Doesn't matter what it is: RP game, action game, or MMO. I just need something to play and get that dopamine rush of getting stuff and progressing. Played a lot of World // Zero, Survive the Night, Royale High, and Island. Any kind of progression system (aside from Tycoons, I suppose). \\ 
\midrule
\textbf{Explanation Provided} & 
This game features a robust progression system through various activities like building and decorating houses, customizing characters, and leveling up skills, similar to "Royale High" and "Island," making it a good fit for your preferences. \\ 
\midrule
\textbf{Game Recommended} & Bloxburg \\ 
\textbf{Game URL} & \url{https://www.xxxx.com/games/185655149} \\ 
\midrule
\textbf{Score Assigned} & 3 \\ 
\midrule
\textbf{Evaluator Comments} & 
The user appears to be seeking player progression in an action/RPG/MMO-style game. While Bloxburg offers a progression system, it is primarily roleplay-oriented, and progression is not a core focus of the experience, making it partially aligned with the user's request. \\ 
\bottomrule
\end{tabular}%
}
\end{table*}
\newpage

\end{document}